\documentclass{article}
\usepackage{axodraw}
\title{\centerline
\bf Gauge invariance of elementary particle processes taking place in
presence of a  background magnetic field}
\bigskip
\author{Kaushik Bhattacharya
\thanks{e-mail
addresses:kaushikb@prl.res.in,
}\\
\normalsize
Physical Research Laboratory, Ahmedabad 380009, India
}
\begin{document}
\newcommand{\Tr}{\mathop{\rm Tr}\nolimits}
\newcommand{\para}{_\parallel}
\newcommand{\pr}{_\perp}
\newcommand{\fs}{\rlap/}
\def\twidle{\widetilde}
\def\f{\frac}
\def\omit#1{_{\!\rlap{$\scriptscriptstyle \backslash$}
{\scriptscriptstyle #1}}}
\def\vec#1{\mathchoice 
	{\mbox{\boldmath $#1$}}
	{\mbox{\boldmath $#1$}}
	{\mbox{\boldmath $\scriptstyle #1$}}
	{\mbox{\boldmath $\scriptscriptstyle #1$}}
}
\def\eqn#1{Eq.\ (\ref{#1})}
\maketitle
\begin{abstract}
Elementary particle scatterings and decays in presence of a background
magnetic field are very common in physics, specially after the
observation that the core of the neutron stars can sustain a magnetic
field of the order of $10^{13}\,{\rm G}$. The important point about
these calculations is that they are done in a background of a gauge
field and as a result the calculations are prone to gauge
arbitrariness. In this work we will investigate how this gauge
arbitrariness is eradicated in processes where the initial and final
particles taking part in the interactions are electrically
neutral. Some comments on those processes where the initial or final
state consists of electrically charged particles is presented at the
end of the article.
\end{abstract}
\section{Introduction}
\label{int}
\begin{figure}[h]
\begin{center}
\begin{picture}(180,75)(-90,-20)
\ArrowLine(80,0)(40,0) 
\Text(60,-10)[b]{$\nu$} 
\ArrowLine(40,0)(-40,0)
\Text(0,-10)[b]{$\ell$} 
\ArrowLine(-40,0)(-80,0)
\Text(-60,-10)[b]{$\nu$} 
\PhotonArc(0,0)(40,0,180){2.5}{10.5}
\ArrowArcn(0,27)(20,120,60)
\Text(0,53)[b]{$W^-$}
\end{picture}
\qquad
\begin{picture}(180,75)(-90,-20)
\ArrowLine(50,0)(0,0) 
\Text(25,-10)[c]{$\nu$} 
\ArrowLine(0,0)(-50,0)
\Text(-25,-10)[c]{$\nu$} 
\Photon(0,0)(0,30)24
\Text(5,15)[l]{$Z$}
\ArrowArc(0,45)(15,-90,270)
\Text(0,67)[]{$\ell$}
\end{picture}
\caption[]{\sf One-loop diagrams for neutrino self-energy.  
\label{f:selfen}}
\end{center}
\end{figure}
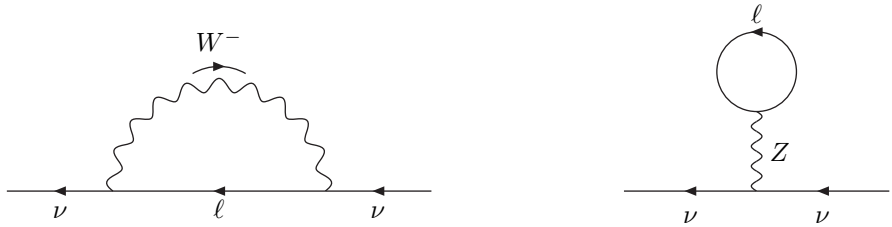

Calculations of elementary particle decays and scattering
cross-sections in presence of a background magnetic field are commonly
found in literature \cite{Can69, MOc69, MOc70, Dorofeev:az}. These
calculations became more important after it was understood that the
neutron star cores can sustain magnetic fields of the order of
$10^{13}\,{\rm G}$ or more. In presence of such strong magnetic fields
all the particles which have a magnetic moment are bound to get
affected and consequently their properties like self-energy, decay
rates or scattering cross-sections are modified. The interesting
feature of a background magnetic field is that it not only modifies
properties of particles with magnetic moments but can also affect the
properties of particles which do not have any magnetic moment. The
obvious question is, how is it possible? To give a satisfactory answer
to this question we take the example of standard model neutrinos and
their self-energy.  In the standard model, neutrinos have no electric
charge hence they do not have any direct coupling to photons in any
renormalizable quantum field theory.  The standard Dirac contribution
to the magnetic moment, which comes from the vector coupling of a
fermion to the photon, is therefore absent for the neutrino. In the
standard model of electroweak interactions, the neutrinos cannot have
any anomalous magnetic moment either. The reason is: anomalous
magnetic moment comes from chirality-flipping interactions
$\bar\psi\sigma_{\mu\nu}\psi F^{\mu\nu}$, and neutrinos cannot have
such interactions because there are no right-chiral neutrinos in the
standard model. Consequently the standard model neutrinos cannot
interact with the background magnetic field.  If one looks at the
Feynman diagrams contributing to the neutrino self-energy in
Fig.~\ref{f:selfen} one will see that the self-energy diagrams contain
charged lepton, $\ell$, propagators and $W^-$ propagator. Although the
neutrinos in this case have no coupling with the background fields the
virtual particles in the loops can have couplings with the background
field and so the result of the self-energy calculations are bound to
be affected by the background fields. 

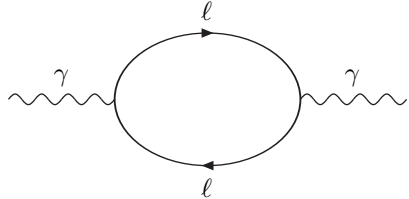
\begin{figure}[btp]
\begin{center}
\begin{picture}(150,50)(0,-25)
\Photon(0,0)(40,0){2}{4}
\Text(20,5)[b]{$\gamma$}
\Photon(110,0)(150,0){2}{4}
\Text(130,5)[b]{$\gamma$}
\Text(75,30)[b]{$\ell$}
\Text(75,-30)[t]{$\ell$}
\Oval(75,0)(25,35)(0)
\ArrowLine(74,25)(76,25)
\ArrowLine(76,-25)(74,-25)
\end{picture}
\end{center}
\caption[One-loop diagram for vacuum polarization.]{One-loop diagram 
for vacuum polarization of the photon.}
\label{f:1loop}
\end{figure}

As a second example we take the case of the vacuum polarization of the
photon. To one-loop the Feynman diagram is shown in
Fig.~\ref{f:1loop}. In this case also the virtual particles
propagating in the loops must be charged particles and as a result the
magnetic field will affect the expression of the vacuum polarization
result.  There can be various other cases, as an example the four
photon interactions in QED, in which the magnetic field can
affect properties of electrically neutral particles.

The calculations of the quantities like the neutrino self-energy and
the vacuum polarization of the photons in presence of a background
magnetic field is similar to the calculations of them in absence of
the same, except that now one has to use the modified two-point functions of
the charged particles. We will discuss about the charged fermion
two-point functions in presence of a magnetic field in section \ref{cfp}.

Before going into further details about how to calculate various
elementary particle processes in presence of a background uniform
magnetic field, we must be careful about the fact that we are
calculating physical quantities in presence of a gauge
field. The background field must be specified by
some suitable choice of gauge. Then the question is whether the
quantities which we calculate are in the end gauge invariant?  If the
answer is no than there is no utility of doing all those calculations
because they will be arbitrary to the gauge which one uses. In this
article we will try to understand how the gauge arbitrariness of the
various calculations mentioned above can be tackled. We will mainly
focus on all those elementary particle processes whose Feynman
diagrams do not have any charged particle external legs. Processes
like Compton scattering, pair annihilation and so on will be dealt
briefly in the penultimate section.

The article is organized in the following way. In section \ref{pm} we
will set the preliminaries required to follow the following sections
and doing so we will also set the mathematical notations which we will
follow.  In section \ref{cfp} we will discuss about the form of the
Schwinger phase accompanying the charged fermion two-point function in
presence of a magnetic field and try to understand its properties. The
next section will deal with the effects of the phase factor on loop
calculations. It will be discussed in which circumstances we can
neglect the phase factor for an actual calculation and when this
cannot be done. In section \ref{cpel} we show the Dirac equation
solutions in a magnetic field background and a pure gauge background
and comment on their gauge transformation properties and the
difficulties we come across to prove background gauge invariance of
those processes which involve charged particles in the external lines
of the Feynman diagrams.
\section{Conventions and notations}
\label{pm}
In presence of a background magnetic field we can decompose the photon 
field as follows:
\begin{eqnarray}
A^\mu(x)=A^\mu_{\rm D}(x) + A^\mu_{\rm B}(x)\,,
\label{da}
\end{eqnarray}
where $A^\mu_{\rm D}(x)$ is the dynamical photon field and $A^\mu_{\rm
B}(x)$ is the classical background field which gives rise to the
magnetic field. If the uniform background classical magnetic field is
called ${\vec B}$ then we must have:
\begin{eqnarray}
{\vec B}=\nabla \times {\vec A}_{\rm B}({\vec x})\,,
\label{mag}
\end{eqnarray}
where $A^\mu_{\rm B}(x)=(0, {\vec A}_{\rm B}({\vec x}))$. In presence
of the background magnetic field we can also write the field strength
tensor as:
\begin{eqnarray}
F^{\mu \nu}(x) = F^{\mu \nu}_{\rm D}(x) + F^{\mu \nu}_{\rm B}\,,
\label{fmunu}
\end{eqnarray}
where $F^{\mu \nu}_{\rm D}(x)=\partial^\mu A^\nu_{\rm D}(x) -
\partial^\nu A^\mu_{\rm D}(x)$ and $F^{i j}_{\rm B}= \partial^i
A^j_{\rm B}(x) - \partial^j A^i_{\rm B}(x)$ is a constant as
given in Eq.~(\ref{mag}).

The QED Lagrangian can be written as:
\begin{eqnarray}
{\mathcal L}=\overline{\psi}(i\gamma_\mu D^\mu - m)\psi -
\frac{1}{4}F^{\mu \nu}F_{\mu \nu}\,,
\label{qed1}
\end{eqnarray}
where $D^\mu=\partial^\mu + ieq A^\mu$ is the covariant derivative of
the fermion fields. Here $e$ is the magnitude of the electronic charge
and $q$ designates the sign, for electrons $q=-1$ and for positrons
$q=+1$. In this article we will only be talking about those cases
where $|q|=1$.  The gauge invariant Lagrangian can also be written as:
\begin{eqnarray}
{\mathcal L}=\overline{\psi}\left[\gamma_\mu \Pi^\mu - m\right]\psi -
e q \overline{\psi}\gamma_\mu\psi A^\mu_{\rm D} - \frac{1}{4}
F^{\mu \nu}F_{\mu \nu}\,,
\label{qed2}
\end{eqnarray}
where $\Pi^\mu=i\partial^\mu - eq A^\mu_{\rm B}$ is the kinetic momentum
of the fermions in presence of the background field. The commutation 
relations of interest are:
\begin{eqnarray}
[x^\mu, \Pi^\nu]=-i g^{\mu \nu}\,,~~~~~~~~{\rm and}~~~~~~~~~[\Pi^\mu,
\Pi^\nu]= -i eq F^{\mu \nu}_{\rm B}\,,
\label{comm}
\end{eqnarray}
where,
\begin{eqnarray}
g^{\mu \nu}={\rm diag}(1, -1, -1, -1)\,.
\label{gmunu}
\end{eqnarray}
Now we can talk about the gauge invariance of the Lagrangian given in
Eq.~(\ref{qed2}). We have to be particularly careful since there are
two kind of gauge invariances here. The first kind is about the gauge
invariance of the Lagrangian under a gauge transformation of the
dynamical photon field $A^\mu_{\rm D}\to A^\mu_{\rm D} + \partial_\mu
\omega$ where $\omega$ is some well behaved function of space and
time. Under this kind of a gauge transformation the fermion fields
will transform like $\psi \to \psi\,e^{-ieq\omega}$. But this does not
exhaust all the gauge transformation possibilities of the Lagrangian
given in Eq.~(\ref{qed2}), we can still have a gauge transformation of
the background gauge field $A^\mu_{\rm B}$ as $A^\mu_{\rm B} \to
A^\mu_{\rm B} + \partial^\mu \lambda$ where $\lambda$ is a well
behaved function of space-time coordinates, which will leave the
Lagrangian in Eq.~(\ref{qed2}) invariant provided the fields
transforms like $\psi \to \psi\,e^{-ieq\lambda}$.

The Euler-Lagrange equation of $\psi$ is:
\begin{eqnarray}
(\fs\Pi - m)\psi(x)=0\,,
\label{dirac}
\end{eqnarray}
where $\fs\Pi=\gamma_\mu \Pi^\mu$. Generally the inverse of the
kinetic part of the Lagrangian (for Eq.~(\ref{qed2}) the quantity
sandwiched between the fields in the first term in the right hand
side of the equation) gives the two-point Greens function of the
theory, which are the building blocks of subsequent developments. In
the present circumstance, the two-point functions are also defined
as:
\begin{eqnarray}
(\fs\Pi - m)S_{\rm B}(x,x')=\delta^4 (x-x')\,,
\label{2ptf}
\end{eqnarray}
which can also be written in a matrix form:
\begin{eqnarray}
S_{\rm B}=\frac{1}{\fs\Pi - m}\,,
\label{gf}
\end{eqnarray}
where $S_{\rm B}$ is a matrix, in fact the inverse matrix
of $\fs\Pi - m$, and the two-point Greens function is $\langle
x'|S_{\rm B}|x\rangle=S_{\rm B}(x,x')$. So from the definition of the
two-point function given in Eq.~(\ref{2ptf}) we see that the fermion
two-point functions are defined not in a gauge invariant way but has
explicit dependence on the background gauge field. We will talk about
this gauge dependence of the two-point function in the next section.

In the Feynman diagrams appearing in Fig.~\ref{f:selfen} and
Fig.~\ref{f:1loop}, the photons represent the dynamical photons
$A^\mu_{\rm D}$ appearing in the Lagrangian given in
Eq.~(\ref{qed2}). The background field $A^\mu_{\rm B}$ is tacitly
taken into care by the fermion propagators or other charged particle
propagators appearing inside the loops.  Another interesting point
which must be noted is that although the Lagrangian as written down in
Eq.~(\ref{qed2}) is Lorentz invariant but it contains $ A^\mu_{\rm B}$
and $A^\mu_{\rm B}$ satisfies Eq.~(\ref{mag}) with the condition that
$A^0_{\rm B}(x)=0$. The condition as given in Eq.~(\ref{mag}) is not
Lorentz invariant i.e., a magnetic field is not a Lorentz invariant
quantity.  So if we demand that we will be working in presence of a
uniform classical background magnetic field then we are restricting
the Lorentz covariance of the theory.

Lastly we fix the notations and conventions which we will be following
in this article. In future the magnitude of the uniform background
magnetic field will always be denoted by the symbol $B$ and we will
take it to be pointed along the $z$-direction of the coordinate axis.
The magnetic field vector pointing in the $z$-direction will be given
by
\begin{eqnarray}
\vec{B} = {\hat z} B\,,
\end{eqnarray}
where ${\hat z}$ is the unit vector along the $z$-axis.

As Lorentz covariance is restricted and the magnetic field chooses a
particular direction in space, the 4-vector structure breaks down into
a perpendicular part and a parallel part. If $a^{\mu}$ is a 4-vector,
then
\begin{eqnarray}
a^{\mu}_\parallel &=& (a^0, 0,  0, a^3)\,, 
\label{defpara}\\
a^{\mu}_\perp &=& (0, a^1,  a^2, 0)\,,
\label{defpr}
\end{eqnarray}
such that
\begin{eqnarray}
a^{\mu} = a^{\mu}_\parallel + a^{\mu}_\perp\,.
\label{fvectordec}
\end{eqnarray}
Also in our convention
\begin{eqnarray}
g_{\mu \nu} = g^{\parallel}_{\mu \nu} + g^{\perp}_{\mu \nu}\,,
\label{Gmunupa}
\end{eqnarray}
where
\begin{eqnarray}
g^{\parallel}_{\mu \nu} & = & (1,0,0,-1)\,,\\
g^{\perp}_{\mu \nu} & = & (0,-1,-1,0)\,.
\end{eqnarray}
Terms as $a^2_{\parallel}$ and
$a^2_{\perp}$ stands for:
\begin{eqnarray}
a^2_\parallel &=& g^{\parallel}_{\mu \nu} a^\mu a^\nu\,,\\
          &=& (a^1)^2  -  (a^3)^2\,,
\label{defparasq}\\
a^2_\perp   &=& - g^{\perp}_{\mu \nu} a^\mu a^\nu\,,\\
          &=&  (a^1)^2 +  (a^2)^2\,,
\label{defprsq}
\end{eqnarray}
such that 
\begin{eqnarray}
a^2 = a_\parallel^2 - a_\perp^2\,.
\end{eqnarray}
Also in our case,
\begin{eqnarray}
F^{12}_{\rm B}=-F^{21}_{\rm B}=B\,.
\end{eqnarray}
With this amount of preliminary understanding about the fermion
two-point functions in a background magnetic field and the notations which we
will follow we move to the next section which deals with the explicit
form of the two-point fermionic Greens functions.
\section{The phase factor appearing in the two-point Greens function of a 
charged fermion}
\label{cfp}
In a much celebrated paper Schwinger derived what will be the form of
the fermionic two-point function in presence of constant
electromagnetic fields \cite{Schwinger:1951nm, Itzuber,
Chyi:1999fc}. In the present article we will only study the behaviour
of the two-point function in presence of a pure magnetic field.  Later
on it was possible to find a curious momentum space description of the
Schwinger's two-point function in presence of a magnetic field. It can
be noted that the above two-point function is not a free field
propagator, it contains interactions with the background magnetic
field. This fact makes the two-point function to be translationally
non-invariant. The reason why it becomes translationally non-invariant
is closely linked to its gauge transformation properties which we will
discuss shortly afterwards in this section. As translational
invariance is lost we cannot ideally Fourier transform $S_{\rm
B}(x,x')$ into something analogous to $S_{\rm B}(p)$, but the actual
two-point function contains a multiplicative factor which is translation
invariant and has all the good properties that a two-point function
should have.

Schwinger's calculation of the fermion two-point function relies on solving
the operator formula given in Eq.~(\ref{gf}) and then finding out the
matrix element $\langle x'|S_{\rm B}|x\rangle=S_{\rm B}(x,x')$. The
two-point function can be expressed as \cite{Chyi:1999fc, Tsai:1974ap, Tsai:1974fa, Bhattacharya:2002aj}:
\begin{eqnarray}
iS_{\rm B}(x,x') = \kappa(x,x') \int \frac{d^4 p}{(2 \pi)^4} e^{-ip \cdot
(x-x')} iS_{\rm B} (p) \,,
\label{schwingprop}
\end{eqnarray}
here $x$ stands for the coordinate 4-vector as usual.  $S_{\rm B}(p)$
is expressed as an integral over a variable $s$, usually (though
confusingly) called the `proper time':
\begin{eqnarray}
i S_{\rm B} (p) =\int_0^\infty ds\; e^{\Phi(p,s)} G(p,s) \,.
\label{SB}
\end{eqnarray}
The quantities $\Phi(p,s)$ and $G(p,s)$ can be written in the
following way :
\begin{eqnarray}
\Phi(p,s) &\equiv& is \left( p_\parallel^2 - {\tan
(eqBs) \over eqBs} \, p_\perp^2 - m^2 \right) - \epsilon s \,, 
\label{Phi}
\\
G(p,s) &\equiv&  {e^{ieBs\Sigma\!_z} \over \cos(eqBs)} \;
\left( 
\rlap/p_\parallel + {e^{-ieqBs\Sigma_z} \over \cos(eBs)}
\rlap/ p_\perp + m \right) \nonumber\\*
&=& ( 1 + i\Sigma_z \tan(eqBs) ) 
(\rlap/p_\parallel + m ) + \sec^2 (eqBs) \rlap/ p_\perp \,.
\label{G}
\end{eqnarray}
In the above expressions $\Sigma_z=i\gamma^1\gamma^2$, where the gamma
matrices are taken to be in the standard Dirac-Pauli representation.
$\fs p_\parallel=g^{\parallel}_{\mu \nu}\gamma^\mu p^\nu$ and $\fs
p_\perp=g^{\perp}_{\mu \nu}\gamma^\mu p^\nu$ while the symbols
$p_\parallel^2$ and $p_\perp^2$ are explained in the last
section. $\epsilon$ is an infinitesimal positive quantity introduced
for the convergence of the integrals. For convenience henceforth we
will call the expression of the two-point function as given in
Eq.~(\ref{schwingprop}) as the Schwinger two-point function.

In a typical loop diagram, one therefore will have to perform not only
integrations over the loop momenta, but also over the proper time
variables.  From the above equations we see that $i S_{\rm B} (p)$ is
manifestly translation invariant, and it has another interesting
property.  As $B \to 0$, $\Phi(p,s) \to is (p_\parallel^2 - p_\perp^2
- m^2) - \epsilon s= is(p^2 -m^2) - \epsilon s$ and $G(p,s)\to
(\rlap/p_\parallel + m )+\rlap/ p_\perp = \fs p +m$. Therefore when $B
\to 0$ we have:
\begin{eqnarray}
i S_{\rm B} (p) &\to& \int_0^\infty ds\; e^{is(p^2 -m^2) - \epsilon s}
(\fs p +m)\,,\\
&=& \frac{i(\fs p + m)}{p^2 - m^2 + i\epsilon}\,,
\label{ffp}
\end{eqnarray} 
which is the normal fermion propagator in absence of any background
magnetic field. More over it is seen that $iS_{\rm B} (p)$ consists of
$B$s and not the gauge fields, so $iS_{\rm B} (p)$ is not only
translation invariant but gauge invariant also. This leads us to the
conclusion that the gauge dependence of the two-point function $iS_{\rm
B}(x,x')$ must present in the function $\kappa(x,x')$.

Not going into any detailed description of $\kappa(x,x')$ we can
simply understand its necessity in the two-point function.  In
presence of a background gauge field the gauge transformation property
of the fields of the charged fermions are different at two different
space-time points. Unless there is some factor in the two-point
function which can connect these two fields at different space-time
points with different gauge transformation properties, the
calculations involving charged fermion two-point functions will not be
manifestly gauge covariant.  The gauge transformed fields comes with
phase factors where the phase depends upon the space-time point where
the gauge transformation is made. The fermionic fields at two
different space-time points will therefore have two different phase
factors. To make a connection between them $\kappa(x,x')$ must also be
some form of a phase.  Conventionally it is named the phase-factor.
The phase factor, as calculated by Schwinger \cite{Schwinger:1951nm},
is given as:
\begin{eqnarray}
\kappa(x,x') = \exp \left\{ ieq I(x,x')\right\}
\label{PF}
\end{eqnarray}
where
\begin{eqnarray}
I(x,x') = \int_{x'}^x d\xi_\mu \left[A^\mu_{\rm B}(\xi) +
\frac12 F^{\mu \nu}_{\rm B} (\xi - x')_\nu\right]\,.
\label{IPF}
\end{eqnarray}
From Eq.~(\ref{IPF}) we notice that the phase
factor breaks the translation invariance of the two-point function.

For a constant background field we can always write the gauge field as
\begin{eqnarray}
A^\mu_{\rm B}(\xi) = - \frac12 F^{\mu \nu}_{\rm B} \xi_\nu  + 
\partial^\mu \lambda(\xi)\,,
\label{ConsF}
\end{eqnarray}
where $\lambda(\xi)$ is an arbitrary well behaved function and depends
upon our choice of gauge.  Using the above relation in conjunction
with Eq.~(\ref{IPF}) we can simplify the integration appearing in the
phase factor as
\begin{eqnarray}
I(x,x') = \int_{x'}^x d\xi_\mu \left[- \frac12
F^{\mu \nu}_{\rm B} x'_\nu + \partial^\mu \lambda(\xi) \right]\,.
\end{eqnarray}
Using the constancy of the field strength tensor the above expression
can be written as
\begin{eqnarray}
I(x,x')= \frac12 x'_\mu F^{\mu \nu}_{\rm B} x_\nu + \lambda(x) - 
\lambda(x')\,. 
\label{compactIPF}
\end{eqnarray}
From Eq.~(\ref{compactIPF}) we can immediately see if we set $x = x'$,
in other words if we integrate over any closed contour in space-time
$I(x,x')$ vanishes. Thus $I(x,x')$ connecting two points in space-time
is independent of the path joining them, and as a result the phase
factor of the Schwinger two-point function joining the points $x'$ and $x$ in
Eq.~(\ref{PF}) is also path independent.

Utilizing the path independence of the phase factor of the two-point function 
the general convention is to choose a straight line path connecting
the two points $x'$ and $x$. Points on this path are represented by
\begin{eqnarray}
\xi^\mu = (1 - \zeta) x'^\mu + \zeta x^\mu\,,
\label{stpath}
\end{eqnarray}
where the parameter $\zeta$ ranges from $0$ to $1$. Using
Eq.~(\ref{IPF}) and the straight line path given in Eq.~(\ref{stpath})
one gets
\begin{eqnarray}
I(x,x') &=& \int_{x'}^x d\xi_\mu A^\mu_{\rm B}(\xi) + \frac{\zeta}{2}
\int_0^1 d\zeta (x_\mu - x'_\mu)F^{\mu \nu}_{\rm B}(x_\nu - x'_\nu)\,,
\nonumber\\
&=&  \int_{x'}^x d\xi_\mu A^\mu_{\rm B}(\xi)\,.
\label{AI}
\end{eqnarray}
Using Eq.~(\ref{ConsF}) for the gauge field we can retrieve
Eq.~(\ref{compactIPF}). 

The form of Eq.~(\ref{PF}) when $I(x,x')$ is as given in
Eq.~(\ref{AI}) is very similar to a Wilson line but it is not a
general Wilson line. A general Wilson line in presence of an U(1)
field is defined as:
\begin{eqnarray}
U(x,x')=\exp\,\left[ieq\int_P d\xi_\mu A^\mu_{\rm B}\right]\,,
\label{WL}
\end{eqnarray}
where $P$ specifies a particular path joining the points $x$ and $x'$
in space-time. From Eq.~(\ref{AI}) we see that in this case $I(x,x')$
is obtained by integrating $A^\mu_{\rm B}(\xi)$ only on a straight
line path joining $x'$ and $x$ as given in Eq.~(\ref{stpath}). For a general Wilson line as defined in Eq.~(\ref{WL}) the gauge invariant Wilson loop will be:
\begin{eqnarray}
U(x,x)=\exp\,\left[ieq\frac{e}{2}\oint_{\cal S} d\sigma_{\mu \nu} 
F^{\mu \nu}_{\rm B}\right]\,,
\label{WLP}
\end{eqnarray}
where ${\cal S}$ is a surface that spans the closed integration loop and
$d\sigma_{\mu \nu}$ is an area element on this surface. As such
$U(x,x')$ will not be unity in general but as the integration defining
$I(x,x')$ in Eq.~(\ref{AI}) is only on a straight line path so in this
case $\kappa(x,x)$ will always be unity.

From Eq.~(\ref{compactIPF}) it is clear that the phase factor is
dependent on the form of the function $\lambda(\xi)$, that is to say
the fermion two-point function is dependent on the gauge in which the constant
background magnetic field is specified. 
Suppose we are working in such a gauge that $\lambda(\xi) = 0$, and
then we make a gauge transformation of the background field as
\begin{eqnarray}
A^\mu_{\rm B} \to A^\mu_{\rm B} + \partial^\mu \lambda(\xi)
\end{eqnarray}
then from Eq.~(\ref{schwingprop}) it follows that the fermion
two-point function will transform as 
\begin{eqnarray}
iS_{\rm B}(x,x') \to \exp(ie\lambda(x))iS_{\rm B}(x,x')\exp(-ie\lambda(x'))\,.
\label{GTP}
\end{eqnarray}
under the gauge transformation. As because we are working in presence
of a background gauge field, the fields of the charged particles and
their two-point functions both become background gauge dependent. This is the
reason why the phase factor arises in the expression of the
two-point function.  

In the $B \to 0$ limit $i S_{\rm B} (p)$ goes to the normal fermionic
propagators in absence of any magnetic field, but what is the fate of
$\kappa(x,x')$ when $B \to 0$? The answer is closely related to the
way one choose $A^\mu_{\rm B}$. There are many equivalent ways of
writing $A^\mu_{\rm B}$ as:
\begin{eqnarray}
A^0_{\rm B} = A^y_{\rm B} = A^z_{\rm B} = 0 \,, \qquad 
A^x_{\rm B} = -yB + b\,.
\label{GA}
\end{eqnarray}
or 
\begin{eqnarray}
A^0_{\rm B} = A^x_{\rm B} = A^z_{\rm B} = 0 \,, \qquad 
A^y_{\rm B} = xB+c\,.
\label{GB}
\end{eqnarray}
or
\begin{eqnarray}
A^0_{\rm B} = A^z_{\rm B} = 0 \,, \qquad 
A^y_{\rm B} = \frac{1}{2} xB+d\,, \qquad
A^x_{\rm B} = -\frac{1}{2} yB+g\,,
\label{GC}
\end{eqnarray}
where $b$, $c$, $d$ and $g$ are constants. In the above equations $x$,
$y$ are just coordinates and not 4-vectors.  All of these above
choices gives a magnetic field along the $z$-axis of the proper
magnitude and the point to note about the above choices of the gauge
fields is that all of them goes to zero as $B \to 0$ if
$b=c=d=g=0$. When $b=c=d=g=0$ and $B \to 0$ it is noted that $iS_{\rm
B} (p)$ approaches the free propagator form as it should be in absence
of the gauge fields and $\kappa(x,x')$ also approaches unity, as is
expected. But $b=c=d=g=0$ is not the most general gauge choice which
produces a magnetic field along the $z$-direction.  With the general
choice of the gauges as is specified in Eq.~(\ref{GA}), Eq.~(\ref{GB})
or Eq.~(\ref{GC}) it is clear that $\kappa(x,x')$ does not approach
unity as $B\to 0$. If we take the choice of the background gauge field
as given in Eq.~(\ref{GA}) then $\kappa(x,x')\to e^{ieqb(x - x')}$ as
$B\to 0$. So when $B \to 0$ in this case we will have,
\begin{eqnarray}
iS_{\rm B}(x,x') \to e^{ieqb(x - x')} \int \frac{d^4 p}{(2 \pi)^4} e^{-ip \cdot
(x-x')}\frac{i(\fs p + m)}{p^2 - m^2 + i\epsilon}\,,
\label{schpropb0}
\end{eqnarray} 
which is translation invariant but not the form which we expect when there 
is no background magnetic field present. 

A some what similar situation arises when we deal with pure gauge
configurations. The Schwinger two-point function as given in
Eq.~(\ref{schwingprop}) is specific for the case of those gauges which
gives a background magnetic field. In the general derivation by
Schwinger~\cite{Schwinger:1951nm} the form of the fermionic two-point function
was derived for general constant $F^{\mu \nu}_{\rm B}$. If we choose a
pure gauge field which does not give rise to any electric or magnetic
field as
\begin{eqnarray}
A^\mu_{\rm B}(\xi) = k \delta^\mu_{\,\,\,\alpha} \xi^\alpha\,,
\label{PG}
\end{eqnarray}
where $k$ is a constant, then also we can find the Schwinger
two-point function and it differs from the vacuum propagator only by the phase
factor. Using Eq.~(\ref{AI}) in this case we will have,
\begin{eqnarray}
iS_{\rm B}(x,x') = e^{(ieqk(x^\alpha x_\alpha - x'^\alpha x'_\alpha))}
             \int \frac{d^4 p}{ 2 \pi^4} e^{-ip \cdot(x-x')} \frac{i(\rlap/p + m)}{p^2 - m^2 + i\epsilon}\,.
\label{PGP}
\end{eqnarray}
This form of the propagator is not translation invariant and does not
match our expected form. The exponential factor multiplieng the
propagator of the free fermions is annoying. In the next section we
will explicitly show how to tackle these problems.

A brief summary of the properties about the Schwinger two-point function is
presented below. The Schwinger two-point function, in general, is a function
of $A^\mu_{B}$ and is defined for all gauge configurations including
pure gauges. In the present section only its form in presence of a
uniform magnetic field was presented.  As a consequence of this fact
when one takes the $B \to 0$ limit of it, the free-fermion propagator
is reproduced up to multiplicative phase, which designates a trivial
pure background gauge configuration.  In other words the $B \to 0$
limit of the two-point function means a transition (not a gauge
transformation) of $iS_{\rm B}(x,x')$ where initially the gauge fields
followed Eq.~(\ref{mag}) with $A^0_{\rm B}(x)=0$ and finally
$A^\mu_{\rm B}(x)=\partial \rho(x)$ for some well behaved function
$\rho(x)$. The gauges which produces a magnetic field along the
$z$-axis may contain constant terms as $b$, $c$, $d$, $g$ and when
$B=0$ all the gauges in Eq.~(\ref{GA}), Eq.~(\ref{GB}) and
Eq.~(\ref{GC}) becomes trivial pure gauge configurations. So in
general the Schwinger two-point function as given in Eq.~(\ref{schwingprop})
can be continuously transformed from the case where it is a function
of the gauge fields which produces a magnetic field to the case where
it is a function of pure gauge fields. This fact has an interesting
outcome. It is known that in presence of an external magnetic field
the transverse momenta $p_\perp$ do not represent gauge invariant
degrees of freedom. But still the expression of the Schwinger
two-point function contains $p^2_\perp$ explicitly. The presence of
$p^2_\perp$ is there only because the two-point function is a function
of $A^\mu_{\rm B}(x)$ and the domain of $A^\mu_{\rm B}(x)$ which can
give rise to a magnetic field can be continuously transformed into
trivial pure gauge configurations.
\section{Effects of $\kappa(x,x')$ on calculations}
\label{loops}
As we have discussed previously in section \ref{cfp}, the phase factor
in the Schwinger two-point function appears because the fermion two-point function
attaches two points with different gauge transformation
properties. Till now we have only talked about fermion two-point functions but
one can also find out the Greens functions of charged scalars and
vector particles in presence of a uniform background magnetic
field. It is seen that all of these two-point functions can be written as a
product of a translationally invariant part and a background gauge
dependent part. In fact all of the gauge dependent parts of the various
two-point functions are same and are equal to $\kappa(x,x')$
\cite{Erdas:1998uu}. The charged gauge boson two-point functions also depend
upon other gauge parameters, like Feynman gauge parameter or Landau
gauge parameter, but those are related to the dynamical gauge
invariance. As we are not interested in an actual loop calculation we
will not explicitly write down the forms of all the two-point functions here
but for further discussions we will only utilize the fact that the
translationally non-invariant part of all the two-point functions of
charged particles are functionally equivalent to $\kappa(x,x')$. So
all the points we will use to prove that a two-point fermion loop is
independent of the choice of the background gauge will also apply for
a loop like the one which is to the left in Fig.~\ref{f:selfen}. 

In this section we will study various cases and try to understand what
will be the effect of $\kappa(x,x')$ on loop calculations.  To
understand its importance we take an example of the one loop photon
vacuum polarization in QED. Let $P$ and $Q$ be the two space-time
coordinates where the photon line interacts with the virtual charged
fermions.  If we are interested in finding out the overall phase
factor accompanying the vacuum polarization tensor then we will have
to use the Schwinger two-point function for the charged fermions. The
contribution from the phase factors $\kappa(Q,P)$ and $\kappa(P,Q)$ to
the loop integral in coordinate space, which we denote as $\Phi(P,Q)$
will be according to Eq.~(\ref{PF}) and Eq.~(\ref{compactIPF})
\begin{eqnarray}
\Phi(P,Q) &=& \kappa(Q,P) \kappa(P,Q)\nonumber\\
           &=& \exp\left\{ ieq\frac12 \left[P_\mu F^{\mu \nu}_{\rm B} Q_\nu +
           Q_\mu F^{\mu \nu}_{\rm B} P_\nu \right]\right\}\,,
\label{VP}
\end{eqnarray}
which reduces to unity because of the antisymmetry of $F^{\mu
\nu}_{\rm B}$. From Eq.~(\ref{VP}) it is seen that the phase factor's
contribution in the one loop calculation is trivial and obviously
gauge invariant.  The same analysis also holds for the diagram in the
left of Fig.~\ref{f:selfen}.

Next we take a loop with three charged particle two-point functions connecting
three space-time points $P,~Q,~R$.  The overall phase contribution to
the loop integral in the coordinate space can then be calculated using
Eq.~(\ref{PF}) and Eq.~(\ref{compactIPF}) and is given by
\begin{eqnarray}
\Phi(P,Q,R) &=& \kappa(Q,P) \kappa(R,Q) \kappa(P,R)\nonumber\\
          &=& \exp\left\{ ieq\frac12 \left[P_\mu F^{\mu \nu}_{\rm B} Q_\nu +
          Q_\mu F^{\mu \nu}_{\rm B} R_\nu + R_\mu F^{\mu \nu}_{\rm B} P_\nu\right]\right\}\,.
\label{TPF}
\end{eqnarray}
As all the phase factors are of the same form the first point to
notice is that the contribution from the function $\lambda(\xi)$
cancel out in the overall factor, showing that the contribution is
explicitly gauge invariant.

The next point which requires to be discussed is about the path
independence of the phase factor. From section \ref{cfp} we know that
$\kappa(Q,P)$ is independent of the path which joins them, but here
the path independence of $\kappa(Q,P)$ does not imply
\begin{eqnarray}
\Phi(P,Q,R) = \kappa(Q,P) \kappa(R,Q) \kappa(P,R) = 1\,,
\label{one}
\end{eqnarray}
or from Eq.~(\ref{compactIPF})
\begin{eqnarray}
I(P,P) = I(Q,P) + I(R,Q) + I(P,R) = 0.
\label{AdIPF}
\end{eqnarray}
Instead of the above expectation we get a finite contribution from
Eq.~(\ref{TPF}). As $I(Q,P)$ consists of the product of the two end
points instead of their difference, the different phase factors from
the different paths connecting two intermediate points of the loop
when multiplied does not reduce to unity. In general if, 
\begin{eqnarray}
I(Q,P)=f(Q)-f(P)\,, 
\label{q-p}
\end{eqnarray}
where $f(P)$ is some well-behaved function of space and time, then
Eq.~(\ref{AdIPF}) will hold and consequently Eq.~(\ref{one}) will be
true.  In this regard we can say that the factor which multiplied the
the normal fermion propagator in absence of any magnetic field in
Eq.~(\ref{schpropb0}) is of the form as given in Eq.~(\ref{q-p}) and
consequently they will not pose any problems for actual loop
calculations.
\subsection{Similarity of $\Phi$ with the Wilson loop}
\label{s:FR}
In the previous discussions it was shown that the phase factor
contribution to the one loop calculations of various cases are
explicitly gauge invariant. In the case of the photon vacuum
polarization it was shown that it contributes nothing for the phase
factor. The contribution from the phase factors for cases where one
has three or more than three vertices can be understood in another
way.  If we have a loop with three or more vertices then the overall
phase will be $\oint_L d\xi^\mu A^\mu_{\rm B}(\xi)$ where $L$
designates the path which joins all the vertices in straight
lines. Then from generalized Stokes theorem we can write:
\begin{eqnarray}
\oint_L d\xi^\mu A^\mu_{\rm B}(\xi) = \frac{1}{2}\int_{\cal S} 
d\sigma_{\mu \nu} 
F^{\mu \nu}_{\rm B}(\xi)\,,
\label{wc}
\end{eqnarray}
where ${\cal S}$ is the area of the loop enclosed by the straight
lines and $d\sigma_{\mu \nu}$ is the infinitesimal surface area in the
$\xi_\mu - \xi_\nu$ plane and $F^{\mu \nu}_{\rm B}$ is the field
strength tensor. In this case we notice that $\kappa(x,x')$ can be
interpreted as the Wilson loop which is a gauge invariant quantity.

For pure gauge configurations we have seen in Eq.~(\ref{PGP}) that the
fermion two-point function in presence of a pure gauge field comes with an
unwanted phase.  In the level of two-point functions this problem remains but
if we look at loop integrals then this problem of the unwanted phase
disappears. For the specific gauge choice in Eq.~(\ref{PG}) we see
that $I(P,Q)$ is of the form as given in Eq.~(\ref{q-p}) and
consequently the overall phase factor will be trivial. But the result
is not dependent on the particular form of the pure gauge chosen.
From Eq.~(\ref{wc}) we can see in all those cases where one calculates
some process which contains three or more than three vertices in
presence of a background pure gauge field $\Phi=1$ because for a pure
gauge configuration $F^{\mu \nu}_{\rm B}(\xi)=0$.

From Eq.~(\ref{wc}) we can generalize that the overall phase depends
on the flux of the magnetic field attached to the area of the
loop. Given an arbitrary diagram we can initially calculate this flux
to find out the overall phase. The typical nature of the overall phase
function $\Phi(P,Q)$ is that $\Phi(P,P)$ is always unity. Similarly
for all those graphs containing only two charged particle two-point functions,
meeting each other at vertices $P$ and $Q$, the overall phase function
$\Phi(P,Q)=1$. But if the number of vertices in the loop, containing
solely charged particles, is three or more then $\Phi$ is non
trivial and it behaves like the Wilson loop as defined in
Eq.~(\ref{WLP}).  
\section{Processes with charged particles as external legs in their 
Feynman diagrams}
\label{cpel}
Except the processes which we have discussed in the previous sections
there are many other processes where there are charged particles in
the external legs of the Feynman diagrams, as Compton scattering,
electron electron-neutrino scattering, pair creation and annihilation
and so on. In addition to the charged external particles all these
processes may require virtual charged particles for their happening.
In this cases the standard procedure is to solve the Dirac equation in
presence of the background magnetic field. But in these cases the
calculations loose manifest gauge invariance because we have to solve
Eq.~(\ref{dirac}) choosing some form of the background gauge and that
will fix the gauge in this case.  To make the point clear we actually
solve the Dirac equation in presence of a background magnetic field in
this section. For stationary states, we can write the solution of the
Dirac equation as:
\begin{eqnarray}
\psi = e^{-iEt} \left( \begin{array}{c} \phi \\ \chi \end{array}
\right) \,,
\end{eqnarray}
where $\phi$ and $\chi$ are 2-component objects.  We use the Pauli-Dirac
representation of the Dirac matrices. 
In this notation, we can write 
\begin{eqnarray}
(E-m)\phi &=& \vec \sigma \cdot (-i\vec\nabla - eq\vec A_{\rm B}) \chi \,, 
\label{eq1}\\*
(E+m) \chi &=& \vec \sigma \cdot (-i\vec\nabla - eq\vec A_{\rm B}) \phi \,,
\label{eq2}
\end{eqnarray}
where $\vec \sigma$ designates the Pauli matrices. Eliminating $\chi$,
we obtain
\begin{eqnarray}
(E^2 - m^2)\phi &=& \Big[ \vec \sigma \cdot (-i\vec\nabla - eq\vec A_{\rm B}) 
\Big]^2 \phi \,.
\label{phieq1} 
\end{eqnarray}
Now we choose the background gauge field
configuration as given in Eq.~(\ref{GA}) with $b=0$ and 
with this choice Eq.\ (\ref{phieq1}) reduces to the form
\begin{eqnarray}
(E^2 - m^2)\phi 
&=& \Big[ -\vec\nabla^2 + (eqB)^2 y^2 - eqB(2iy
{\partial\over 
\partial x} + \sigma_z) \Big] \phi \,. 
\label{phieq} 
\end{eqnarray}
Here $\sigma_z$ is the diagonal Pauli matrix.  With this choice of our
background gauge field the positive energy solutions of the Dirac
equation for an electron is given as:
\begin{eqnarray}
e^{-ip\cdot X {\omit y}} U_s (y,n,\vec p \omit y) \,,
\end{eqnarray}
here $X^\mu$ denotes the space-time coordinate and $p\cdot X {\omit
y}= E_n t - \vec p\cdot\vec X{\omit y} = E_n t - p_x x+p_z z$. Here we
have introduced the notation $\vec X$ for the spatial coordinates (in
order to distinguish it from $x$, which is one of the components of
$\vec X$), and $\vec X\omit y$ for the vector $\vec X$ with its
$y$-component set equal to zero. In other words, $\vec p\cdot \vec
X{\omit y} \equiv p_xx+p_zz$, where $p_x$ and $p_z$ denote the
eigenvalues of momentum in the $x$ and $z$ directions.\footnote{It is
to be understood that whenever we write the spatial component of any
vector with a lettered subscript, it would imply the corresponding
contravariant component of the relevant 4-vector.} The energy $E_n$ is
given as:
\begin{eqnarray}
E_n^2 = m^2 + p_z^2 + 2neB \,,
\label{En}
\end{eqnarray}
giving the relativistic form of Landau energy levels where $n$ is a 
natural number. The form of $U_s$ for $s=+1,-1$ are given by \cite{Bhattacharya:2002qf},
\begin{eqnarray}
U_+ (y,n,\vec p \omit y) = \left( \begin{array}{c} 
I_{n-1}(\xi) \\[2ex] 0 \\[2ex] 
{\strut\textstyle p_z \over \strut\textstyle E_n+m} I_{n-1}(\xi) \\[2ex]
-\, {\strut\textstyle \sqrt{2neB} \over \strut\textstyle 
E_n+m} I_n (\xi) 
\end{array} \right) \,, \qquad 
U_- (y,n,\vec p \omit y) = \left( \begin{array}{c} 
0 \\[2ex] I_n (\xi) \\[2ex]
-\, {\strut\textstyle \sqrt{2neB} \over \strut\textstyle E_n+m}
I_{n-1}(\xi) \\[2ex] 
-\,{\strut\textstyle p_z \over \strut\textstyle E_n+m} I_n(\xi) 
\end{array} \right) \,. 
\label{Usoln}
\end{eqnarray}
In the above expressions 
\begin{eqnarray}
\xi = \sqrt{eB} \left( y + {p_x \over eqB}
\right) \,, 
\end{eqnarray}
and 
\begin{eqnarray}
I_{\nu}(\xi)=\left( {\sqrt{eB} \over \nu! \, 2^\nu \sqrt{\pi}} \,
\right)^{1/2}\, e^{-\xi^2/2} H_{\nu}(\xi) \,,
\end{eqnarray}
where $\nu$ are also natural numbers, and $H_{\nu}(\xi)$ are Hermite 
polynomial functions of order $\nu$.  

The negative energy solutions are:
\begin{eqnarray}
e^{ip\cdot X{\omit y}} V_s (y,n, \vec p\omit y) \,,
\end{eqnarray}
where 
\begin{eqnarray}
V_+ (y,n,\vec p\omit y) = \left( \begin{array}{c} 
{\strut\textstyle p_z \over \strut\textstyle E_n+m}
I_{n-1}(\widetilde\xi) \\[2ex] 
{\strut\textstyle \sqrt{2neB} \over \strut\textstyle E_n+m} 
I_n (\widetilde\xi)  \\[2ex] 
I_{n-1}(\widetilde\xi) \\[2ex] 0
\end{array} \right) \,, \qquad 
V_- (y,n,\vec p\omit y) = \left( \begin{array}{c} 
{\strut\textstyle \sqrt{2neB} \over \strut\textstyle E_n+m}
I_{n-1}(\widetilde\xi) \\[2ex] 
-\,{\strut\textstyle p_z \over \strut\textstyle E_n+m}
I_n(\widetilde\xi)  \\[2ex] 
0 \\[2ex] I_n (\widetilde\xi)
\end{array} \right) \,.
\label{Vsoln}
\end{eqnarray}
where $\widetilde\xi$ is obtained from $\xi$ by changing the sign of
the $p_x$-term. The calculation of the above solutions are given in
the appendix at the end.

The above solutions are obtained by solving the Dirac equation for a
particular background gauge field configuration as specified in
Eq.~(\ref{GA}) with $b=0$.  If we designate the Dirac solutions in a
background magnetic field by $\psi_B(x)$ then the above solution is
one of them but we can easily gauge transform this solution to obtain
other solutions. All the gauge configurations we have specified in
Eq.~(\ref{GA}), Eq.~(\ref{GB}) and Eq.~(\ref{GC}) can be connected by
smooth gauge transformations. Specifically if we take $b=0$ and $c=0$
then we can get the gauge fields in Eq.~(\ref{GB}) from those of
Eq.~(\ref{GA}) by the transformation $A^\mu_{\rm B} \to A^\mu_{\rm B}
+ \partial^\mu \lambda$ where $\lambda(x,y)=xyB$. In this case
$\psi'_B(x)=\psi_B(x) \,e^{-ieqxyB}$ where $\psi_B(x)$ is the solution
for the gauge configuration as given in Eq.~(\ref{GA}) and
$\psi'_B(x)$ is the solution of the Dirac equation for the gauge
configuration as given in Eq.~(\ref{GB}) for $b=c=0$. In the above
solutions the transverse momenta, $p_x$ and $p_y$ are just spurious
degrees of freedom, they depend on the choice of the gauge. In the
above solutions we will not find $p_y$ as the gauge we chose to work
with contained $y$, but had we started with the gauge fields as given
in Eq.~(\ref{GB}) then $p_x$ should have been absent and if we had
worked with the gauge fields as specified in Eq.~(\ref{GC}) none of
$p_x$ or $p_y$ should have appeared in our calculations. The
arbitrariness of $p_x$ and $p_y$ reflects the fact that in presence of
electromagnetic gauge fields $p_x$ and $p_y$ are not the proper
quantities to work with. But in what ever background gauge we choose
to work with the expression of the energy as given in Eq.~(\ref{En}),
which is a physically measurable gauge invariant quantity, remains the
same.

It can be a curious exercise to see if a pure gauge configuration like
$A^0_{\rm B}=A^y_{\rm B}=A^z_{\rm B}=0$ and $A^x_{\rm B}=xk$, where
$k$ is a constant, can produce any effect on the Dirac solutions. We
saw in our previous discussion with the Schwinger two-point functions that a
pure gauge configuration modifies its form but cannot affect loop
calculations involving them. In the present case also it can be shown
that pure gauge configurations although modifies the shape of the
Dirac solutions but the free-particle solutions can be easily
distinguished from the complete solutions. With the gauge which we
have chosen we can proceed similarly to the case where we were solving
the Dirac equation in a magnetic field, the equation corresponding to
Eq.~(\ref{phieq}) in this case will be:
\begin{eqnarray}
(E^2 - m^2)\phi 
&=& \Big[\left(-i\frac{\partial}{\partial x} -eqxk\right)^2 -
\frac{\partial^2}{\partial y^2} - \frac{\partial^2}{\partial z^2} 
\Big] \phi \,. 
\label{nphieq} 
\end{eqnarray}
Looking at the above equation we can propose a possible solution as:
\begin{eqnarray}
\phi = e^{i \vec {\scriptstyle p} \cdot \vec{\scriptstyle X} \omit x}
f(x) \,, 
\label{nphiform}
\end{eqnarray}
where $f(x)$ is a 2-component matrix which depends only on the
$x$-coordinate. In the present case $\vec p\cdot \vec X{\omit x}
\equiv p_y y+p_zz$.  Putting the above solution into
Eq.~(\ref{nphieq}) we get the differential equation for $f(x)$ which
looks like:
\begin{eqnarray}
\left(-i\frac{d}{dx}- eqxk\right)^2 f(x)= (E^2 - m^2 - p^2_y - p^2_z) f(x)\,.
\label{pgauge}
\end{eqnarray}
Now we can always assume a solution of the above equation of the form:
\begin{eqnarray}
f(x)=e^{ix(a + \frac{1}{2}eqkx)}\eta_\pm\,,
\label{pgsoln}
\end{eqnarray}
where $a$ is a constant and $\eta_\pm$ are the standard two-component 
spinors of the form
\begin{eqnarray}
\eta_+ = \left( \begin{array}{c} 1 \\ 0 \end{array} \right) \,,
\qquad 
\eta_- = \left( \begin{array}{c} 0 \\ 1 \end{array} \right) \,.
\end{eqnarray}
Now if we put this solution into Eq.~(\ref{pgauge}) to get
\begin{eqnarray}
(E^2 - m^2 - p_y^2 - p_z^2 - a^2)f(x)=0\,, 
\end{eqnarray}
which predicts $E^2 - m^2 - p_y^2 - p_z^2 - a^2=0$ and is the
free-particle energy-momentum relation if we assume the constant $a$
to be the momentum canonically conjugate to $x$, i.e. if $a=p_x$.
So Eq.~(\ref{nphiform}) can now be written as:
\begin{eqnarray}
\phi = e^{i \vec {\scriptstyle p} \cdot \vec{\scriptstyle x} } 
e^{\frac{1}{2}ieqkx^2}\eta_\pm \,,
\end{eqnarray}
which is similar to the free-particle solution up to the phase
$e^{ieqkx^2}$. Using the expression of $\phi$ we can find $\chi$ from
Eq.~(\ref{eq2}). It can be seen easily that the complete solution of
the Dirac equation in the pure gauge which we chose is the
free-particle solution times a phase.  The above solution can also be
interpreted in another way, all the solutions of the Dirac equation in
presence of a pure gauge field are related to the free-particle
solutions (solutions with $A^\mu_{\rm B}=0$) by a gauge rotation. In
the present case the rotation phase is simply
$e^{\frac{1}{2}ieqkx^2}$. In presence of a magnetic field the Dirac
solutions show that motion of the electrons transverse to the magnetic
field directions gets quantized yielding discrit eigenvalues as
$n$. This fact is crucial in making the $B \to 0$ limit non trivial in
the present circumstances.

The solutions as given in Eq.~(\ref{Usoln}) and Eq.~(\ref{Vsoln}) can
be used to write the Dirac field in the usual sense and quantize the
system in presence of a magnetic field \cite{Das:1995bn,
Kobayashi:1983dt}. Then using those fields one can define a two-point
function of the fermions. But as the Dirac solutions which we obtained
are solved for a particular gauge field configuration the calculations
with them will not be manifestly gauge invariant but reflect the
choice of the gauge. This situation is unlike the one which we faced
when we were talking about the Schwinger two-point functions where
gauge invariance of the loops can be manifestly proved.  It may be
tempting to write:
\begin{eqnarray}
iS_{\rm B}(x,x')= \langle \Omega|T\{\Psi_B(x)\,\overline{\Psi}_B(x')\}
|\Omega\rangle\,,
\label{to}
\end{eqnarray}
where $\Psi_B(x)$ is the Dirac field in presence of a background
magnetic field and $|\Omega\rangle$ is the appropriate vacuum state
here, it is not the same as free-field vacuum as now the vacuum
contains an infinite number of photons. $iS_{\rm B}(x,x')$ is the
Schwinger two-point function as defined in Eq.~(\ref{schwingprop}). If
we analyze the left and the right hand sides of the above equation we
will see that such a relation cannot be correct. From the discussion
presented in the end of section \ref{cfp} it was observed that the
Schwinger two-point function in the right is a function of $A^\mu_{B}$
and is defined for all gauge configurations including pure gauges and
consequently it had dependence on the transverse momenta components
$p_\perp$ and had a smooth $B\to 0$ limit.  On the other hand
$\Psi_B(x)$ which appears on the right hand side of Eq.~(\ref{to}) is
made up of Dirac solutions in a particular gauge which only follows
Eq.~(\ref{mag}) and as a result of this the $p_\perp$ components are
spurious. There is no transformation which transforms
Eq.~(\ref{Usoln}) or Eq.~(\ref{Vsoln}) to the pure gauge solutions as
discussed in the previous paragraph. Precisely, the right hand side of
Eq.~(\ref{to}) is not a function of a general $A^\mu_{\rm B}(x)$ and
this fact makes it different from the left hand side. As a consequence
it is impossible to obtain the Schwinger two-point function by
manipulating on the Dirac solutions which we obtain by solving Dirac
equation in a magnetic field.

From the discussions in this section we see that it is impossible to
prove generally the gauge invariance of all those elementary particle
processes which includes charged particles as external lines. The way
of proving it must be particular and will depend upon the process
under consideration.
\section{Conclusion}
In this article we tried to analyze the background gauge invariance of
elementary particle processes in presence of a uniform background
magnetic field. It was shown that the background gauge invariance of
the calculations of certain class of processes requires a term in the
fermion two-point functionss, in presence of a magnetic field, which
must have proper covarince properties. In loops solely composed of
charged particles the Schwinger phase will not contribute until the
loop has three or more than three vertices. It was shown that pure
gauge configurations will not alter any properties of the calculations
and $B\to 0$ limit of the calculations yield the expected results.
All the results stated are to one-loop but it is expected that the
result is more general and can be prooved for higher loop cases also
in a way similar to the one presented in this article.  Lastly we
discuss all those particular processes which has one or more charged
particles as external lines in their Feynman diagrams. In this cases
it becomes impossible to furnish a general proof which will show that
all these processes are background gauge invariant, as the
calculations heavily rely on the effect of the particular gauge chosen
to solve the Dirac equation in presence of a background magnetic
field. The only way left to prove these processes to be background
gauge invariant is to calculate a process in multiple gauges and show
that the result is the same.
\vskip .5cm
{\bf Acknowledgement:} Many topics discussed in the present article
took shape while discussing with Palash Baran Pal and Prashant
Panigrahi. The author acknowledges their contribution in shaping the
ideas presented.
\appendix\section*{\hfil Appendix \hfil}
\section{Solution of the Dirac equation in a constant background 
magnetic field}
\label{app}
In the standard Dirac-Pauli representation:
\begin{eqnarray}
\vec \alpha = \left( \begin{array}{cc} 
		0 & \vec\sigma \\
		\vec\sigma & 0
	      \end{array} \right) \,, \qquad
\beta = \left( \begin{array}{cc} 
		1 & 0 \\
		0 & -1 
	      \end{array} \right)
\label{}
\end{eqnarray}
where each block represents a $2\times2$ matrix.  Noticing that the
coordinates-ordinates $x$ and $z$ do not appear in Eq.~(\ref{phieq}) except
through the derivatives a possible solution of it can be:
\begin{eqnarray}
\phi = e^{i \vec {\scriptstyle p} \cdot \vec{\scriptstyle X} \omit y}
f(y) \,, 
\label{phiform}
\end{eqnarray}
where $f(y)$ is a 2-component matrix which depends only on the
$y$-coordinate, and possibly some momentum components, as we will see
shortly.  There will be two independent solutions for $f(y)$, which
can be taken, without any loss of generality, to be the eigenstates of
$\sigma_z$ with eigenvalues $s=\pm 1$. This means that we choose the
two independent solutions in the form
\begin{eqnarray}
f_+ (y) = \left( \begin{array}{c} F_+(y) \\ 0 \end{array} \right) \,,
\qquad 
f_- (y) = \left( \begin{array}{c} 0 \\ F_-(y) \end{array} \right) \,.
\end{eqnarray}
Since $\sigma_z f_s = sf_s$, the differential equations satisfied by
$F_s$ is
\begin{eqnarray}
{d^2F_s \over dy^2} - (eqBy + p_x)^2 F_s + (E^2 - m^2 -
p_z^2 + eqBs) 
F_s = 0 \,,
\label{Fseqn}
\end{eqnarray}
which is obtained from Eq.\ (\ref{phieq}).  The solution is obtained
by using the dimensionless variable
\begin{eqnarray}
\xi = \sqrt{e |q| B} \left( y + {p_x \over eqB}
\right) \,, 
\label{xi}
\end{eqnarray}
which transforms Eq.\ (\ref{Fseqn}) to the form
\begin{eqnarray}
\left[ {d^2 \over d\xi^2} -\xi^2 + a_s \right] F_s = 0 \,,
\end{eqnarray}
where
\begin{eqnarray}
a_s = {E^2 - m^2 - p_z^2 + eqBs \over e|q|B} \,.
\end{eqnarray}
This is a special form of Hermite's equation, and the solutions exist
provided $a_s=2\nu+1$ for $\nu=0,1,2,\cdots$. This provides the energy
eigenvalues 
\begin{eqnarray}
E^2 = m^2 + p_z^2 + (2\nu+1)e|q|B - eqBs \,,
\label{E}
\end{eqnarray}
and the solutions for $F_s$ are
\begin{eqnarray}
N_{\nu} e^{-\xi^2/2} H_{\nu}(\xi) \equiv I_{\nu}(\xi) \,,
\label{In}
\end{eqnarray}
where $H_\nu$ are Hermite polynomials of order $\nu$, and $N_\nu$ are
normalizations which we take to be
\begin{eqnarray}
N_\nu = 
\left( {\sqrt{e|q|B} \over \nu! \, 2^\nu \sqrt{\pi}} \,
\right)^{1/2} \,.  
\label{Nn}
\end{eqnarray}
With our choice of normalization, the functions $I_\nu$
satisfy the completeness relation
\begin{eqnarray}
\sum_\nu I_\nu(\xi) I_\nu(\xi_\star) = \sqrt{e|q|B} \;
\delta(\xi-\xi_\star) = \delta (y-y_\star) \,, 
\label{completeness}
\end{eqnarray}
where $\xi_\star$ is obtained by replacing $y$ by $y_\star$ in Eq.\
(\ref{xi}).

So far, $q$ was arbitrary.  We now specialize to the case of
electrons, for which $q=-1$.  The solutions are then conveniently
classified by the energy eigenvalues
\begin{eqnarray}
E_n^2 = m^2 + p_z^2 + 2neB \,,
\end{eqnarray}
which is the relativistic form of Landau energy levels. The solutions
are two fold degenerate in general: for $s=1$, $\nu=n-1$ and for
$s=-1$, $\nu=n$.  In the case of $n=0$, only the second solution is
available since $\nu$ cannot be negative.  The solutions can have
positive or negative energies. We will denote the positive square root
of the right side by $E_n$. Representing the solution corresponding to
this $n$-th Landau level by a superscript $n$, we can then write for
the positive energy solutions,
\begin{eqnarray}
f_+^{(n)} (y) = \left( \begin{array}{c} 
I_{n-1}(\xi) \\ 0 \end{array} \right) \,,
\qquad 
f_-^{(n)} (y) = \left( \begin{array}{c} 
0 \\ I_n (\xi) \end{array} \right) \,.
\label{fsolns}
\end{eqnarray}
For $n=0$, the solution $f_+$ does not exist. We will consistently
incorporate this fact by defining
\begin{eqnarray}
I_{-1} (y) = 0 \,,
\label{I_-1}
\end{eqnarray}
in addition to the definition of $I_n$ in Eq.\ (\ref{In}) for
non-negative integers $n$.

The solutions in Eq.\ (\ref{fsolns}) determine the upper components of
the spinors through Eq.\ (\ref{phiform}). The lower
components, denoted by $\chi$ earlier, can be solved using
Eq.\ (\ref{eq2}).


\begin{thebibliography}{999}

\bibitem{Can69} L. Fassio-Canuto, 
Phys. Rev. {\bf 187}, 2141 (1969).

\bibitem{MOc69} J.~J. Matese and R.~F. O'Connell, 
Phys. Rev. {\bf 180}, 1289 (1969). 

\bibitem{MOc70} J.~J. Matese and R.~F. O'Connell, 
Astroph. Jour. {\bf 160}, 451 (1970).

\bibitem{Dorofeev:az}
O.~F.~Dorofeev, V.~N.~Rodionov and I.~M.~Ternov,
JETP Lett.\  {\bf 40}, 917 (1984)
[Pisma Zh.\ Eksp.\ Teor.\ Fiz.\  {\bf 40}, 159 (1984)].

\bibitem{Schwinger:1951nm}
J.~S.~Schwinger,
Phys.\ Rev.\  {\bf 82}, 664 (1951).

\bibitem{Itzuber}
C.~Itzykson and J.~Zuber, {\it Quantum Field Theory}, Mc Graw Hill press, 
International student edition, Secon edition, page {\bf 100}.

\bibitem{Chyi:1999fc}
T.~K.~Chyi, C.~W.~Hwang, W.~F.~Kao, G.~L.~Lin, K.~W.~Ng and J.~J.~Tseng,
Phys.\ Rev.\ D {\bf 62}, 105014 (2000)
[arXiv:hep-th/9912134].

\bibitem{Tsai:1974ap}
W.~y.~Tsai,
Phys.\ Rev.\ D {\bf 10}, 2699 (1974).

\bibitem{Tsai:1974fa}
W.~y.~Tsai and T.~Erber,
Phys.\ Rev.\ D {\bf 10}, 492 (1974).

\bibitem{Bhattacharya:2002aj}
K.~Bhattacharya and P.~B.~Pal,
Proc.\,Ind.\,Natl.\,Sci.\,Acad.\, {\bf 70}: 145 (2004) 
arXiv:hep-ph/0212118.

\bibitem{Erdas:1998uu}
A.~Erdas, C.~W.~Kim and T.~H.~Lee,
Phys.\ Rev.\ D {\bf 58}, 085016 (1998)
[arXiv:hep-ph/9804318].

\bibitem{Bhattacharya:2002qf}
K.~Bhattacharya and P.~B.~Pal,
Pramana {\bf 62}, 1041 (2004)
[arXiv:hep-ph/0209053].

\bibitem{Das:1995bn}
A.~K.~Das and M.~Hott,
Phys.\ Rev.\ D {\bf 53}, 2252 (1996)
[arXiv:hep-th/9504086].

\bibitem{Kobayashi:1983dt}
M.~Kobayashi and M.~Sakamoto,
Prog.\ Theor.\ Phys.\  {\bf 70}, 1375 (1983).

\end{thebibliography}
\end{document}